\documentclass[prl,preprint,twocolumn,tightenlines]{revtex4}
\usepackage{amsmath,amssymb,graphicx,bm}

\begin{document}

\preprint{CPHT-RR024.0503}

\title{
Exclusive two $\rho ^0$-mesons production 
in $\gamma\gamma^*$ collision}

\author{I.V. Anikin}
\affiliation{CPHT de l'{\'E}cole Polytechnique,
             91128 Palaiseau Cedex, France\footnote{Unit{\'e} Mixte de 
             Recherche du CNRS (UMR 7644)} and \\
             Bogoliubov Laboratory of Theoretical Physics, 
             JINR, 141980 Dubna, Russia}
\author{B. Pire}
\affiliation{CPHT de l'{\'E}cole Polytechnique,
             91128 Palaiseau Cedex, France\footnote{Unit{\'e} Mixte de 
             Recherche du CNRS (UMR 7644)}}
\author{O.V. Teryaev}
\affiliation{Bogoliubov Laboratory of Theoretical Physics, 
             JINR, 141980 Dubna, Russia}
\vspace{1.5cm}

\begin{abstract}
We give a theoretical description of the L3 Collaboration data 
on the cross-section of the exclusive two $\rho$-mesons production in
$\gamma\gamma ^*$ collision with large photon virtuality. 
These data prove the scaling behaviour of the leading twist 
generalized distribution  amplitude of vector mesons. They thus
prove the relevance of a partonic description of the exclusive
process $\gamma\gamma ^*\to\rho ^0\rho ^0$ at 
$Q^2\geq 1.2\, {\rm GeV^2}$ and $W\leq 3.0 \, {\rm GeV}.$
\vspace{1pc}
\end{abstract}
\maketitle

\section{Introduction}

Two-photon collisions provide a tool
to study a variety of fundamental aspects of QCD and have long been a
subject of great interest (cf, e.g., \cite{Terazawa,Bud,photon-conf}
and references therein). A peculiar facet of this interesting domain is
exclusive two hadron production
in the region where one initial photon is highly virtual (its
virtuality being denoted as $-Q^2$) but the overall energy (or invariant
mass of the two hadrons) is small~\cite{DGPT}. This
process factorizes~\cite{MRG,Freund} into a perturbatively calculable,
short-distance dominated scattering $\gamma^* \gamma \to q \bar q$ or
$\gamma^* \gamma \to g g$, and non-perturbative matrix elements
measuring the transitions $q \bar q \to A B$ and $g g \to  A B $. These
matrix elements  have been called generalized distribution
amplitudes (GDAs) to emphasize their close connection to the
distribution amplitudes introduced many years ago in the QCD
description of exclusive hard processes~\cite{LepageBrodsky}.

In this paper, we focus on the process $\gamma\gamma ^* \to \rho^0 \rho^0$
which has recently been observed at LEP in the right kinematical
domain \cite{L3Coll}. We calculate the differential cross-section of 
the considered processes as a function of $Q^2$
and compare it to the experimental data.

\section{ Kinematics }

The reaction which we study here is:
\begin{eqnarray}
\label{pr}
&&e(k)+e(l)\to
\\ 
&&e(k^{\prime})+e(l^{\prime})+\rho ^0(p_1)+\rho ^0(p_2)
\nonumber
\end{eqnarray}
where the initial  electron $e(k)$ radiates
a hard virtual photon with momentum $q=k-k^{\prime}$~;
in other words, the square of virtual photon momentum $q^2=-Q^2$ is
very large. This means that the scattered electron  $e(k^{\prime})$
is tagged.
To describe reaction (\ref{pr}), it is useful to consider, at the same
time,  the sub-process of reaction (\ref{pr})~:
\begin{eqnarray}
\label{spr}
e(k)+\gamma(q^{\prime})\to e(k^{\prime})+\rho ^0(p_1)+\rho ^0(p_2).
\end{eqnarray}
Regarding the other photon momentum $q^{\prime}=l-l^{\prime}$,
we assume that, firstly, its momentum is collinear to the electron
momentum $l$ and, secondly, that $q^{\prime \, 2}$ is approximately
equal to zero, which is a usual approximation when the second lepton
is untagged.

Let us now pass to a short discussion of kinematics in
the $\gamma\gamma ^*$ center of mass system. We adopt the
$z$ axis directed along the three-dimensional vector ${\bf q}$,
and the $\rho$-meson momenta are in the $(x,z)$-plane.
So, we write for momenta in the c.m. system~:
\begin{eqnarray}
\label{kin}
&&q=(q_0,\,0,\,0,\,{\bf q}),
\nonumber\\
&& p_1=(p_1^0,\, {\bf p}_1{\rm sin}\theta,\,0,\,
{\bf p}_1{\rm cos}\theta)
\end{eqnarray}
Also, we need to write down the Mandelstam $S$-variables for
the electron-positron (\ref{pr}) and electron-photon (\ref{spr})
collisions:
\begin{eqnarray}
S_{ee}=(k+l)^2, \quad S_{e\gamma}=(k+q^{\prime})^2.
\end{eqnarray}
Neglecting the lepton masses, these variables
can be rewritten as
\begin{eqnarray}
S_{ee}\approx 2(k\cdot l), \quad
S_{e\gamma}\approx 2(k\cdot q^{\prime})=x_2 S_{ee},
\end{eqnarray}
where the fraction $x_2$ defined as  $q_0^{\prime}=x_2 l_0 $ is introduced
(see, for instance \cite{Diehl00}).

\section{Parameterization of $\rho$-matrix elements and
their properties }

Let us first introduce the basis light-cone vectors.
We adopt a light-cone basis
consisting of two light-like vectors $p$ and $n$ 
of mass dimension $1$ and $-1$, respectively.
In other words, they obey the following conditions~:
\begin{eqnarray}
\label{lcb}
p^2=n^2=0, \quad (p\cdot n)=1.
\end{eqnarray}
In this basis,
the $\rho$-mesons momenta $p_1 $ and $p_2 $ can be written as
\begin{eqnarray}
\label{lcd1}
&&p_1^{\mu}=\frac{1+\tilde\zeta}{2}p^{\mu}+
\frac{1-\tilde\zeta}{2}\frac{W^2}{2} n^{\mu}-\frac{\Delta_{T}^{\mu}}{2},
\nonumber\\
&&p_2^{\mu}=\frac{1-\tilde\zeta}{2}p^{\mu}+
\frac{1+\tilde\zeta}{2}\frac{W^2}{2} n^{\mu}+\frac{\Delta_{T}^{\mu}}{2}
\end{eqnarray}
As usual, we introduce the sum and difference of hadronic momenta:
\begin{eqnarray}
\label{lcd2}
\Delta^{\mu}=p_2^{\mu}-p_1^{\mu}, \quad
P^{\mu}=p_2^{\mu}+p_1^{\mu}.
\end{eqnarray}
The skewedness parameter $\tilde\zeta$ is defined by
\begin{eqnarray}
\Delta\cdot n =-\tilde\zeta P\cdot n,
\end{eqnarray}
or
\begin{eqnarray}
\frac{1+\tilde\zeta}{2}=\frac{1+\beta\cos{\theta}}{2},
\beta=\sqrt{1-\frac{4m_{\rho}^2}{W^2}}.
\end{eqnarray}
Notice that the relation between our definition and definition 
in, for instance, \cite{Diehl00} reads
\begin{eqnarray}
\frac{1+\tilde\zeta}{2}=\zeta.
\end{eqnarray}
Within this frame the transverse component of the
transfer momentum $\Delta_T=(0,{\bf \Delta}_T,0)$ is given by
\begin{eqnarray}
\Delta_T^2=-{\bf \Delta}_T^2=
\biggl( W^2-4m_{\rho}^2\biggr) \sin^2{\theta}.
\end{eqnarray}

We now come to the parameterization
of the relevant matrix elements.  Keeping
the terms of leading  twist $2$, the vector
and axial correlators can be written 
as (see, for 
instance, \cite{BCDP}, where the crossed
case of the generalized parton distributions in a spin-$1$
particles is considered):
\begin{eqnarray}
\label{par1}
&&\langle p_1,\lambda_1;p_2,\lambda_2 |
\bar\psi(0)\gamma_{\mu} \psi(\lambda n)
| 0 \rangle
\stackrel{{\cal F}}{=}
\\
&&p_{\mu}\sum_{i}
e_1^{\alpha}e_2^{\beta}
V^{(i)}_{\alpha\beta}(p_1,p_2,n)
H^{\rho\rho,\,V}_i(y,\tilde\zeta,W^2),
\nonumber\\
\label{par2}
&&\langle p_1,\lambda_1;p_2,\lambda_2 |
\bar\psi(0)\gamma_5 \gamma_{\mu} \psi(\lambda n)
| 0 \rangle
\stackrel{{\cal F}}{=}
\\
&&p_{\mu}\sum_{i}e_1^{\alpha}e_2^{\beta}
A^{(i)}_{\alpha\beta}(p_1,p_2,n)
H^{\rho\rho,\,A}_i(y,\tilde\zeta,W^2).
\nonumber
\end{eqnarray}
Here $\lambda_1$ and $\lambda_2$ are the helicities of
spin-$1$ hadrons and $\stackrel{{\cal F}}{=}$
denotes the Fourier transformation
with measure ($z_1=\lambda n, z_2=0$) \cite{Ani}:
\begin{eqnarray}
d\mu(y)=dy \,e^{ -iy\,pz_1+i(1-y)\,pz_2}.
\end{eqnarray}
In (\ref{par1}) and (\ref{par2}), the tensor structures 
$V^{(i)}_{\alpha\beta}(p_1,p_2,n)$ and 
$A^{(i)}_{\alpha\beta}(p_1,p_2,n)$ depend on the vectors
$p_1$, $p_2$ and $n$. Notice that due to the parity invariance 
the tensors $V^{(i)}_{\alpha\beta}$ may be written in terms of five
tensor structures while the tensors $A^{(i)}_{\alpha\beta}$
are linear combinations of  four independent structures.

\section{Amplitude of $\gamma\gamma^*\to\rho^0\rho^0$ subprocess}

In this section, we pass to the consideration of
$\gamma(q^\prime)\gamma^*(q)\to\rho^0(p_1)\rho^0(p_2)$ subprocess.
Following \cite{Ani}, the amplitude of this subprocess including
the leading twist-$2$ terms can be written as
\begin{eqnarray}
\label{amp_gg}
&&T_{\mu\nu} =
\int\limits_{0}^{1}
dy\Biggl[
g_{\mu\nu}^T E_-(y)
{\bf V}(y,\cos{\theta},W^2) +
\Biggr.
\nonumber\\
\Biggl.
&&\epsilon^T_{\mu\nu} E_+(y) {\bf A}(y,\cos{\theta},W^2)
\Biggr],
\end{eqnarray}
where 
\begin{eqnarray}
E_\pm=\frac{1}{y}\pm\frac{1}{1-y}.
\end{eqnarray}
In (\ref{amp_gg}), the scalar and pseudo-scalar 
functions (${\bf V}$, ${\bf A}$)
denote the following contraction
\begin{eqnarray}
&&{\bf V}(y,\cos{\theta},W^2)=
\\
&&\sum_i e_1^{\alpha}e_2^{\beta}  V^{(i)}_{\alpha\beta}
H^{\rho\rho,\,V}_i(y,\tilde\zeta(\cos{\theta}),W^2),
\nonumber\\
&&{\bf A}(y,\cos{\theta},W^2)=
\\
&&\sum_i e_1^{\alpha}e_2^{\beta} A^{(i)}_{\alpha\beta}
H^{\rho\rho,\,A}_i(y,\tilde\zeta(\cos{\theta}),W^2).
\nonumber
\end{eqnarray}

The next objects of our consideration are the helicity amplitudes that
are obtained from the usual amplitudes after  multiplying by
the photon polarization vectors
\begin{eqnarray}
\label{hel}
A_{(i,j)}=\varepsilon^{\,\prime\,(i)}_{\mu} \varepsilon^{(j)}_{\nu}
T^{\mu\nu}.
\end{eqnarray}
Here, in the $\gamma\gamma ^*$ c.m. frame system, 
the photon polarization vectors read~
\begin{eqnarray}
\label{pol_vec}
&&\varepsilon^{\,\prime\,(\pm)}_{\mu}=
\left( 0,\frac{\mp 1}{\sqrt{2}},\frac{+i}{\sqrt{2}},0 \right),
\nonumber\\
&&\varepsilon^{\,(\pm)}_{\mu}=
\left( 0,\frac{\mp 1}{\sqrt{2}},\frac{-i}{\sqrt{2}},0 \right), 
\nonumber\\
&&\varepsilon^{\,(0)}_{\mu}=
\left(\frac{|q|}{\sqrt{Q^2}},0,0,\frac{q_0}{\sqrt{Q^2}} \right),
\end{eqnarray}
for the real and virtual photons, respectively.

\section{Differential cross-sections}

We will now concentrate on the calculation of the differential
cross-section of (\ref{pr}). Using the equivalent photon approximation
we find the expression for the corresponding cross-section~:
\begin{eqnarray}
\label{xsec5}
&&\frac{d\sigma}{dQ^2}(e^+e ^-\to e^+ e^-\rho^0\rho^0 )=
\nonumber\\
&&\int..\int dW^2\, d{\rm cos}\theta\, d\phi\, dx_2
\nonumber\\
&&\frac{\alpha}{\pi} F_{WW}(x_2)
\frac{d\sigma(e^+\gamma\to e^+\rho^0\rho^0 )}
{ dW^2\, d{\rm cos}\theta\, d\phi},
\end{eqnarray}
where the Weizsacker-Williams function $F_{WW}$ is defined 
as usual~: 
\begin{eqnarray}
\label{FWW}
&&F_{WW}(x_2)=\frac{1+(1-x_2)^2}{2x_{2}} 
{\rm ln}\frac{Q^{\prime\,2}(x_2)}{m_e^2}
\nonumber\\
&&-\frac{1-x_2}{x_{2}},
\end{eqnarray} 
and the value $Q^{\prime\, 2}$ is defined as 
\begin{eqnarray}
\label{Qpr}
Q^{\prime\, 2}_{max}=-q^{\prime\, 2}_{max}=
(1-x_2)E_2^2 {\rm sin}^2\, \alpha_{max}.
\end{eqnarray}
The angle $\alpha_{max}$ in (\ref{Qpr}) is 
determined by the acceptance of a lepton in the detector 
(see, for instance, \cite{Diehl00}) and the value of 
$\sqrt{S_{ee}}$ is  
$91\, {\rm GeV}$ at LEP1 and $195\, {\rm GeV}$ at LEP2.

In (\ref{xsec5}), the cross-section for the subprocess can be calculated
directly~; we have
\begin{eqnarray}
\label{xsec6}
&&\frac{d\sigma(e^+\gamma\to e^+\rho^0\rho^0 )}
{ dW^2\, d{\rm cos}\theta\, d\phi}=
\frac{\alpha ^3}{16\pi}
\frac{\beta}{S_{e\gamma}^2}\,
\frac{1}{Q^2}
\nonumber\\
&&\Biggl(
1-\frac{2S_{e\gamma}(Q^2+W^2-S_{e\gamma})}
{(Q^2+W^2)^2}
\Biggr) 
\nonumber\\
&& \left| A_{(+,+)}(\cos{\theta},W^2) \right|^2
\end{eqnarray}
where
\begin{eqnarray}
\label{App}
&&\left| A_{(+,+)}(\cos{\theta},W^2) \right|^2=
\\
&&4 \left(
\left|  {\bf V}(\cos{\theta},W^2) \right|^2 +
\left|  {\bf A}(\cos{\theta},W^2) \right|^2
\right).
\nonumber
\end{eqnarray}
In (\ref{App}), the squared and polarization summed
functions $|{\bf V}|^2$ and $|{\bf A}|^2$ read~:
\begin{eqnarray}
\label{V2}
&&|{\bf V}(\cos{\theta},W^2)|^2=
 P^{\alpha_1\alpha_2}(p_1)P^{\beta_1\beta_2}(p_2)
\\
&&\sum_i V^{(i)}_{\alpha_1\beta_1}\int dy_1 E_-(y_1) 
H^{\rho\rho,\,V}_i(y_1,\cos{\theta},W^2)
\nonumber\\
&&\sum_j V^{(j)}_{\alpha_2\beta_2}\int dy_2 E_-(y_2)  
H^{\rho\rho,\,V}_j(y_2,\cos{\theta},W^2)
\nonumber\\
&&|{\bf A}(\cos{\theta},W^2)|^2= 
P^{\alpha_1\alpha_2}(p_1)P^{\beta_1\beta_2}(p_2)
\\
&&\sum_i A^{(i)}_{\alpha_1\beta_1}\int dy_1 E_+(y_1)  
H^{\rho\rho,\,A}_i(y_1,\cos{\theta},W^2)
\nonumber\\
&&\sum_j A^{(j)}_{\alpha_2\beta_2}\int dy_2 E_+(y_2) 
H^{\rho\rho,\,A}_j(y_2,\cos{\theta},W^2),
\nonumber
\end{eqnarray}
where   
\begin{eqnarray}
P_{\alpha\beta}(p)=\sum_{\lambda}
e^{(\lambda)}_{\alpha} e^{*\,(\lambda)}_{\beta}=
-g_{\alpha\beta}+\frac{p_{\alpha}p_{\beta}}{m^2_{\rho}}.
\end{eqnarray} 
The helicity amplitude 
after the integration over $\cos{\theta}$ is implemented
may be written as
\begin{eqnarray}
&&F_{(+,+)}(W^2)=\int d{\rm cos}\theta
\nonumber\\
&&\left| A_{(+,+)}(\cos{\theta},W^2) \right|^2.
\end{eqnarray}
So, the cross-section takes the form
\begin{eqnarray}
\label{xsec7}
&&\frac{d\sigma}{dQ^2}(e^+e ^-\to e^+ e^-\rho^0\rho^0 )=
\nonumber\\
&&\frac{\alpha^4}{16\pi^2} \int dx_2 F_{WW}(x_{2})
\Biggl(
\Biggr.
\nonumber\\
\Biggl.
&&\frac{1}{16 x_2^2 E_2^4 Q^2}
\int dW^2 \beta \,
F_{(+,+)}(W^2)-
\Biggr.
\nonumber\\
\Biggl.
&&\frac{1}{2 x_2 E_2^2 Q^2}
\int dW^2
\frac{\beta \, F_{(+,+)}(W^2)}
{Q^2+W^2}+
\Biggr.
\nonumber\\
\Biggl.
&&\frac{2}{Q^2}
\int dW^2
\frac{\beta \, F_{(+,+)}(W^2)}
{(Q^2+W^2)^2}
\Biggr).
\end{eqnarray}
Notice that the integrated amplitude $F_{(+,+)}$ is independent
of $Q^2$ up to logarithms. 
This is quite natural for the cases where the factorization theorem 
is applied. Besides, the exact $W^2$ dependence of this
amplitude remain unknown unless some modeling is used. 
However, the mean value theorem gives the possibility
to reduce the three different integrals over $W^2$ in (\ref{xsec7}) to
one integration. Indeed, for our case the mean value theorem
reads~:
\begin{eqnarray}
\label{mvt1}
&&\int dW^2
\frac{\beta \, F_{(+,+)}(W^2)}
{Q^2+W^2}=
\\
&&\frac{1}{Q^2+ \langle W_{1}\rangle ^2}
\int dW^2
\beta \, F_{(+,+)}(W^2),
\nonumber\\
\label{mvt2}
&&\int dW^2
\frac{\beta \, F_{(+,+)}(W^2)}
{(Q^2+W^2)^2}=
\\
&&\frac{1}{(Q^2+\langle W_{2}\rangle ^2)^2}
\int dW^2
\beta \, F_{(+,+)}(W^2)
\nonumber
\end{eqnarray}
with two phenomenological parameters
$\langle W_{1}\rangle $ and $\langle W_{2}\rangle$.
Because of that the values of these parameters, generally speaking,
have the same order of magnitude as $Q$, {\it i.e.} are 
not much less than $Q$, we need to keep 
$\langle W \rangle$-parameters in the prefactors of (\ref{mvt1})
and (\ref{mvt2}). 
Thus, we can see from (\ref{mvt1}) and (\ref{mvt2}) that it is 
useful to 
introduce a third phenomenogical parameter as
\begin{eqnarray}
C_1=\int\limits_{W^2_{min}}^{W^2_{max}} dW^2
\beta \, F_{(+,+)}(W^2),
\end{eqnarray}
where the integration over $W^2$ runs into the limits dictated by 
the experiment. 

Finally, the cross-section (\ref{xsec7}) is  expressed through
these three phenomenological parameters in the following simple way:
\begin{eqnarray}
\label{xsec8}
&&\frac{d\sigma}{dQ^2}(e^+e ^-\to e^+ e^-\rho^0\rho^0 )=
\frac{\alpha^4}{16\pi^2}\, C_1 \, \int dx_2 
\nonumber\\
&&F_{WW}(x_{2})\Biggl(
\frac{1}{16 x_2^2 E_2^4 Q^2}+
\frac{2}{Q^2 (Q^2+\langle W_{2}\rangle ^2)^2}
\Biggr.
\nonumber\\
\Biggl.
&&-\frac{1}{2 x_2 E_2^2 Q^2 (Q^2+\langle W_{1}\rangle ^2)}
\Biggr).
\end{eqnarray}
Within the analysis implemented by the L3 Collaboration, 
the value $W$ is in the interval
$1.1\, {\rm GeV} < W < 3.0\, {\rm GeV}$.
Hence we are able to conclude that the phenomenological parameters
$\langle W_{1} \rangle$ and $\langle W_{2}\rangle$ may take,
generally speaking, any values inside this interval.

\section{Numerical results and Discussion}

In the previous section we obtained a simple expression for
the cross-section as a function of three parameters which are
$\langle W_{1}\rangle$, $\langle W_{2}\rangle$ and $C_1$.
Let us now make a fit of these phenomenological parameters
in order to get the best description of experimental data.
The best values of the parameters
can be found by the method of least squares, $\chi^2$-method,
which flows from the maximum likelihood theorem (see, for instance,
\cite{Mey}).
As usual, the $\chi^2$-sum as a function of parameters is written
in the form~:
\begin{eqnarray}
\label{chi2}
\chi^2=\sum_{i=1}^{N}\Biggl(
\frac{\sigma ^{exp}_{i} - \sigma ^{th}_i({\bf P})}{\delta\sigma_i}
\Biggr)^2,
\end{eqnarray}
where ${\bf P}=\{\langle W_{1}\rangle,\,\langle W_{2}\rangle,\, C_1\}$ 
denotes the set of fitted
parameters~; $\sigma ^{exp}_{i}$ and $\sigma ^{th}_i$ are
the experimental measurements of the cross-section and its
theoretical estimations~; $\delta\sigma_i$ are the statistical errors.
The experimental data for the cross-section of the exclusive
double $\rho^0$ production were taken from the measurement
of the L3 collaboration at LEP \cite{L3Coll}.
Further, minimizing $\chi^2$-sum in (\ref{chi2}) with respect to the
parameters ${\bf P}$ we find that the set of solutions ${\bf P}_{min}$
with the confidence intervals are the following~:
\begin{eqnarray}
\label{sol_chi2}
&&C_1=1.0 \pm 0.12 \, {\rm GeV}^2,
\nonumber\\ 
&&\langle W_2 \rangle =1.2 \pm 0.08\, {\rm GeV},
\nonumber\\
&&\langle W_1 \rangle =2.9 \pm 1.8 \, {\rm GeV}.
\end{eqnarray}
With this the magnitude of $\chi^2$ is equal to $1.25$ and, therefore,
we have 
\begin{eqnarray}
\frac{\chi^2}{{\rm degree\, of \, freedom}}=0.25 < 1
\end{eqnarray}
The confidence intervals were defined for the case of one-standard
deviation. The graphics with the experimental data and theoretical
cross-section for the fitted parameters (\ref{sol_chi2}) are
drawn on Fig. 1.
The obtained value of the normalization constant $C_1$ of order
one is probably related to the energy momentum conservation, which 
should define the scale of GDA, like in two pions \cite{Pol} case.  
Besides, notice that within L3 experiment the interval of
changing for $W$ in the $\gamma\gamma^*$ center of mass is
fixed as $1.1\, {\rm GeV} < W < 3.0\, {\rm GeV}$. Therefore,
we are able to note, taking into account (\ref{sol_chi2}), that
the confidence intervals for parameters $\langle W_1\rangle$ lie 
into the whole
available interval for $W$. In other words, we can infer that
the theoretical cross-section depends on $\langle W_1\rangle$ slightly.
Indeed, despite all the three terms in (\ref{xsec8}) are
participating in our analysis, the behaviour of the
cross-section is mainly dictated by the third term in (\ref{xsec8}).

\begin{figure}[htb]
\includegraphics[width=16pc]{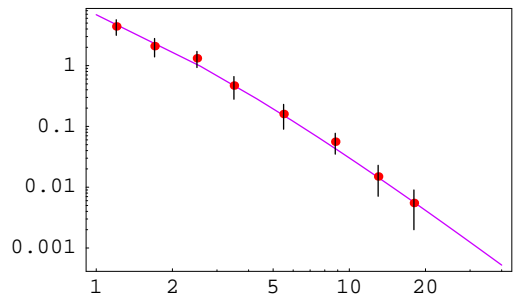}
\caption{Cross-section $d\sigma/dQ^2 [pb/GeV^2]$ as a function
of $Q^2$ .
The theoretical cross-section is plotted for the best fitted
parameters which are $C_1=1.0\, {\rm GeV}^2$ , 
$\langle W_{1} \rangle =2.9 \, {\rm GeV}$ and 
$\langle W_{2}\rangle =1.2 \, {\rm GeV}$.
Experimental data \cite{L3Coll} come from LEP1
and LEP2 runs.}
\end{figure}

\section{Conclusion}

Data thus are in full agreement with the theoretical expectation
on the $Q^2$ behaviour of the cross section. Because of that
the effective structure function $F_{(+,+)}$ is independent
of $Q^2$ up to the logarithms, as it was said before, our analysis 
of the data is a direct proof of the relevance of the partonic
description of the process $\gamma\gamma ^*\to\rho ^0\rho ^0$ 
in the kinematics of the L3 experiment.

Much more can be done if
detailed experimental results are collected. For instance the angular
dependence of the final state is a good test of the validity of the
asymptotic form of the generalized distribution
amplitudes. The spin structure of the final
state, if elucidated, would allow to disentangle the roles of the nine
generalized distribution amplitudes. The $W^2$ behaviour of the cross
section may have some interesting features. It depends much on the
possible resonances which are able to couple to two $\rho$ mesons.
Its Fourier transform allows to have
access to an impact picture of exclusive fragmentation \cite{PS}. 
The $\rho^+ \rho^-$ channel may be calculated along the same lines. In
that case a brehmstrahlung subprocess where the mesons are radiated from
the lepton line must be added. 
The charge and spin asymmetries then come from the
interference of the two processes. In that way one may also study 
the large contribution of $4 \pi$ continuum.  
These items will be discussed in a 
forthcoming publication.

\vskip 0.5cm
We acknowledge useful discussions and correspondance with M. Diehl,
A. Nesterenko and I. Vorobiev.This work has been supported in part 
by RFFI Grant 03-02-16816 and by INTAS Grant (Project 587, call 2000).


\begin{thebibliography}{99}

\bibitem{Terazawa} H.~Terazawa, Rev.\ Mod.\ Phys.\ {\bf 45}
(1973) 615.

\bibitem{Bud} V.M. Budnev {\it et al.}, Phys.\ Rept.\ {\bf 15C}
(1975) 181.

\bibitem{photon-conf} S.J.~Brodsky, hep-ph/9708345, talk presented at
PHOTON~97, Egmond aan Zee, Netherlands, May 1997; \\
M.R.~Pennington, Nucl.\ Phys.\ {\bf B} \ (Proc.\ Suppl.) {\bf 82}
(2000) 291 [hep-ph/9907353].

\bibitem{DGPT} M. Diehl, T. Gousset, B. Pire and O.V. Teryaev, Phys.\
Rev.\ Lett.\ {\bf 81} (1998) 1782 [hep-ph/9805380]; \\
M. Diehl, T. Gousset and B. Pire, Nucl.\ Phys.\ {\bf B} \ (Proc.\
Suppl.) {\bf 82} (2000) 322 [hep-ph/9907453].

\bibitem{MRG} D. Mueller {\it et al.}, Fortschr.\ Phys.\ {\bf 42}
 (1994) 101 [hep-ph/9812448].

\bibitem{Freund} A. Freund, Phys.\ Rev.\ {\bf D61} (2000) 074010
[hep-ph/9903489].

\bibitem{LepageBrodsky} G.P. Lepage and S.J. Brodsky, Phys. Rev. {\bf
D22} (1980) 2157.

\bibitem{L3Coll} L3 Collaboration,  talk  presented 
by S. Nesterov in "PHOTON~03" Workshop, April, 7-11 2003, Frascati, Italy
(these proceedings) 

\bibitem{Diehl00}
M.~Diehl, T.~Gousset and B.~Pire, Phys.\ Rev.\ D {\bf 62} (2000) 073014

\bibitem{BCDP}
E.~R.~Berger, F.~Cano, M.~Diehl and B.~Pire,
Phys.\ Rev.\ Lett.\  {\bf 87} (2001) 142302

\bibitem{Ani}
I.~V.~Anikin and O.~V.~Teryaev,
Phys.\ Lett.\ B {\bf 509} (2001) 95.

\bibitem{Mey} S. L. Meyer, {\it "Data analysis for scientists and
engineers"}, Wiley series in probability and mathematical statistics,
Edt. R. A. Bradley and J. S. Hunter, (1975)

\bibitem{Pol}
M.~V.~Polyakov, Nucl.\ Phys.\ B {\bf 555} (1999) 231
[hep-ph/9809483].

\bibitem{PS}
B.~Pire and L.~Szymanowski, Phys.\ Lett.\ B {\bf 556} (2003) 129
[hep-ph/0212296].

\end{thebibliography}
\end{document}